# Do Consumers Accept AIs as Moral Compliance Agents?

**Associate Professor Greg Nyilasy**
Department of Management and Marketing
Faculty of Business and Economics,
University of Melbourne
198 Berkeley St, University of Melbourne, VIC 3053, Australia

**Abraham Ryan Ade Putra Hito**
Department of Management and Marketing
Faculty of Business and Economics,
University of Melbourne
198 Berkeley St, University of Melbourne, VIC 3053, Australia

Professor Jennifer Overbeck
Melbourne Business School
University of Melbourne
200 Leicester St, Carlton, VIC 3053, Australia

Professor Brock Bastian
School of Psychological Sciences
Faculty of Medicine, Dentistry, and Health Sciences
University of Melbourne, Parkville, VIC 3010, Australia

Professor Darren W. Dahl
UBC Sauder School of Business
University of British Columbia
2053 Mail Mall, Vancouver, BC V6T 1Z2, Canada

# Do Consumers Accept AIs as Moral Compliance Agents?

## Abstract

Consumers are generally resistant to Artificial Intelligence (AI) involvement in moral decision-making, perceiving moral agency as requiring uniquely human traits. This research investigates whether consumers might instead accept AIs in the role of moral compliance, where AI upholds pre-existing moral norms without exercising subjective discretion. Across five studies this research shows that consumers evaluate AI more positively than human agents in moral compliance roles. The findings reveal that this preference arises from inferences of AI's lack of ulterior motives, which are often attributed to human agents. While previous studies have focused on AI as a decision-maker, this work demonstrates the critical role of upholding pre-existing rules, a role in which AI is perceived to excel. These findings contribute to understanding consumer acceptance of moral AI and provide actionable insights for organizations seeking to leverage AI in ethical oversight. By positioning AI as a moral compliance agent, companies can address consumer skepticism, enhance trust, and improve perceptions of corporate ethicality.

# Introduction

Prior work suggests that consumers generally regard AI as ill-suited for the moral domain (Bonnefon et al., 2016). Moral decisions are often high-stakes, rare, and contextually complex, with consequences for fairness, discrimination, well-being, and even survival (Bigman & Gray, 2018; Longoni et al., 2019). People tend to see AI as a logical, outcome-focused system, whereas morality is viewed as requiring a distinctly human mind capable of subjective understanding, ethical awareness, and compassion (K. Gray et al., 2012; Dietvorst & Bartels, 2022). This perceived mismatch between AI's capabilities and the demands of moral judgment fosters avoidance of AI in morally charged decisions, as illustrated by concerns over autonomous vehicles making life-or-death trade-offs between human lives (Bonnefon et al., 2016; Castelo et al., 2019; Clegg et al., 2024).

But what if AI is cast in a different role? What if, instead of being a moral arbiter, making decisions, its moral role is restricted to *moral compliance*, i.e., the mere upholding of moral rules and virtues for decisions somebody else has already made (Geva, 2000; Geva, 2006; Julmi, 2024; Shirai & Watanabe, 2024; Sonenshein, 2007). Could it be that AI is ethically more acceptable, or perhaps even preferred, in a moral compliance role? We suggest that this may be the case as human agents are prone to biases, conflicts of interests, and lapses in moral behavior (Ratner & Miller, 2001). AI, placed in the role of an impartial arbiter, has the promise to overcome such human biases and questionable practices (Bigman et al., 2023; Moon & Kahlor, 2025; Pataranutaporn et al., 2023).

Corporations, non-profit organizations, and governments have already started applying AI in moral compliance roles (OECD, 2025). For example, financial institutions use AIs to uphold moral compliance in sensitive areas such as fraud, money laundering, and rogue internal transactions (Fliderman, 2023; Oracle, 2024). Meta used AI tools to detect online misinformation, signaling consistency and impartiality to stakeholders with these practices (Meta, 2020). Governments and NGOs also see promise in using AI to ensure compliance with ethical procurement, anti-corruption, and ethical finance principles and regulations (Skadden, 2024; United_Nations_Development_Programme, 2024).

AI systems are also used in supply-chain due diligence and in public-procurement integrity. AI solutions can screen such data continuously, apply the same pre-specified rules to every case, and log each decision for audit, thereby signaling impartiality and consistency to stakeholders (Mila_[Quebec_Artificial_Intelligence_Institute], 2025; World_Bank, 2020). By contrast, human compliance officers may be more susceptible to conflicts of interest and to situational pressures from sourcing or sales, which is the very ulterior motive inference our studies document. Our theorized advantage for AI is therefore procedural rather than substantive: AI does not decide whether forced labor is wrong; it enforces the already-agreed rule against forced labor more impartially, more consistently, and with better traceability than conflicted humans typically can.

On a practical level, unpacking these differing perceptions of AI in moral judgement vs. moral compliance roles makes several important contributions. First, public and consumer responses in morally charged contexts function as legitimacy judgments about the ethical acceptability of AI's authority. Understanding these judgments matters because they shape the social license of governance practices and the perceived due process afforded to affected stakeholders. Second, identifying contexts in which people favor AI as a compliance agent provides guidance for ethical institutional design, specifying guardrails (e.g., transparency, recourse, responsibility assignment) under which AI can strengthen moral accountability in governance and regulatory settings rather than displacing it.

From a theoretical perspective, our paper contributes to ongoing AI dialogues in business ethics in three ways. First, we offer a distinct perspective by examining AI in a moral compliance role rather than as a moral decision-maker (Geva, 2000, 2006; Julmi, 2024; Sonenshein, 2007). Prior work has emphasized decision-making "edge cases" and high-stakes dilemmas (e.g., autonomous-vehicle trade-offs; Gill, 2020), yet many morally relevant acts are routine applications of already-settled norms rather than novel adjudications (Bauman et al., 2014; Tooke & Ickes, 1988). By theorizing compliance as the focus of inquiry, we speak directly to legitimacy, due process, and loci of responsibility, clarifying who is answerable when enforcement is delegated to AI and when such delegation is ethically acceptable – to our knowledge, this is the first treatment of this compliance form of AI morality in behavioral contexts. Second, we complement AI–task domain fit research (Castelo et al., 2019) by specifying a task-and-institution fit for algorithmic compliance: Enforcing pre-specified norms demands impartiality, consistency, and traceability. We show how capacities often attributed to AI map onto these ethical requirements. Third, we illuminate why AI may be acceptable in this role: a perceived lack of ulterior motives (Carlson et al., 2022). Across five studies, we connect these mechanisms to downstream judgments about marketers and marketplace actors, positioning consumer responses as ethical legitimacy assessments of governance practices rather than merely instrumental reactions.

The rest of paper is organized as follows. First, we build our theoretical account of why consumers favor AI over humans in moral compliance settings. Second, we present the empirical package, which includes main effect studies (Studies 1 and 2), process evidence (Studies 3 and 4), and a final study that directly contrasts moral compliance and moral decision-making. We discuss these novel findings in the context of relevant ongoing dialogues (theoretical and practical) and point to possible future avenues for research.

## Theoretical Background

### Human Perceptions of AI Moral Decision-making

Prior work suggests that consumers generally perceive AI as a poor fit for the moral domain, especially when the technology is construed as a moral decision-making agent (Bonnefon et al., 2016). Many consumers perceive AI as a logical and analytical entity that seeks to generate the best possible outcomes (Dietvorst & Bartels, 2022). In contrast, the argument is often made that individuals see morality as a domain that requires the decision-maker to possess humanistic subjectivity, a human mind capable of exercising ethical awareness, ethical calculation and reasoning, and compassion (Gray et al., 2012). This can create a perceived mismatch between AI's capabilities and the task domain in question (morality), leading to the avoidance of AI as a suitable decision-making agent in that domain (Castelo et al., 2019; Dietvorst & Bartels, 2022). For instance, the adoption of fully autonomous vehicles (AVs) has been slow due to people's discontent with the possible moral decisions such non-human entities will have to make (Bonnefon et al., 2016). In the event of a traffic accident involving pedestrians, the prospect of having AI systems in autonomous vehicles quantify the value of human lives and make life-or-death decisions is unsettling to many (Clegg et al., 2024).

The high-stakes nature of certain morally relevant situations also discourages reliance on AI for moral decision-making. Some moral choices mean significant and far-reaching ramifications that can deeply alter human lives. The results of a moral decision, thus, can be a matter of fairness, discrimination, well-being, and even life or death (Bigman & Gray, 2018). Hence, despite advancements in AI technology, the gravity of possible consequences may make consumers reluctant to entrust AI with moral judgments. Consumers remain skeptical of AI's ability to match human-level moral reasoning and ability to understand nuanced contexts of morally relevant situations (Dietvorst & Bartels, 2022). In addition, moral dilemmas tend to be novel and to occur rarely. This may lead to the perception that AI lacks precedent to draw on when assessing these unique situations, weighing options, and making decisions (Longoni et al., 2019). Further, in high-stakes decisions, accountability is highly desired. When a moral incident occurs, humans want to be able to question, seek explanations, and ultimately hold someone responsible (Malle et al., 2014). Unfortunately, given AI's stochastic and black-box nature, finding explanations becomes challenging when AI is the decision-maker (Bonezzi et al., 2022). In sum, humans seem reluctant to welcome the prospect of having AI as a moral decision-maker.

**Moral Compliance Fit: AIs vs. Humans**

Nevertheless, there is a less salient but possibly positive perceived advantage to AI's mechanistic rigidity. AI may be accepted in a different role than moral decision-making, in the role of moral *compliance*, i.e., the mere upholding of moral rules and virtues for decisions somebody else has already made (Geva, 2000; Geva, 2006; Julmi, 2024; Shirai & Watanabe, 2024; Sonenshein, 2007). Many everyday moral scenarios are of this nature and do not represent deeply involving decision-dilemmas – instead, they are rather mundane applications of existing rules (Bauman et al., 2014).

Upholding moral compliance, or the "adherence to the moral rules and virtues" (Shirai & Watanabe, 2024, p. 2), is normatively different from adjudicating moral dilemmas (Geva, 2000, 2006; Julmi, 2024; Sonenshein, 2007). Compliance requires consistent rule application, visible neutrality, and credible accountability (Tyler, 1990). Emerging evidence suggests that people sometimes view algorithmic systems as less driven by ulterior motives and more impartial/consistent than humans (Pataranutaporn et al., 2023), which can bolster perceived legitimacy, though acceptance is contingent on transparency and governance safeguards (Martin & Waldman, 2023; Giroux et al., 2022; Lanz et al., 2024; Shah et al., 2024; Pethig & Kroenung, 2023). Building on this literature, we theorize that AIs can be accepted as a moral compliance agent when human conflicts of interest are salient and impartiality is paramount.

Operationally, moral compliance in business ethics is the routinized application of already-established norms to concrete organizational behaviors (Julmi, 2024). For example, imagine a garment manufacturing firm wanting to assure ethical governance in Corporate Social Responsibility (CSR) and sustainable procurement applications (e.g., avoiding forced/child labor in their supply chain). The first step is to commit to external standards such as the *UN Guiding Principles on Business and Human Rights* (United_Nations, 2011) and the *OECD Due Diligence Guidance* for garments and footwear (OECD, 2018), which prohibit forced and child labor, require safe working conditions, and increasingly mandate environmental safeguards. Then comes the application of these standards, which typically requires five tasks: (i) normative literacy (knowing labor/human rights/environmental standards), (ii) supply-chain context knowledge (where subcontracting and other high-risk practices occur), (iii) data/analytics competence to check supplier disclosures, audit reports or external signals, (iv) insulation from conflicts of interest, and (v) documentation/escalation for audit. This is the concrete work implied by our labels "rule application, visible neutrality, and credible accountability" above, as conceptually distinct from originating new moral judgments.

Turning to the role of AI, consumers may be right that AI represents risks in moral decision-making given its current limited capabilities, such as lack of mind and emotion (Gray et al., 2012). However, AI's excellent abilities in processing large amounts of information and reliability in following rules may make it invaluable in complying with moral decisions already made (Arkan et al., 2023; Bonaccio & Dalal, 2010; Sonenshein, 2007). Further, moral rules and norms are social contracts, shaped over centuries among humans, resulting from the need to survive and to live together in a society (Greene, 2014). Thus, one may only need to consult and follow existing moral rules and norms when participating in many morally charged contexts, rather than "re-inventing the moral wheel." For example, in clinical trials and other medical interventions, an ethics review committee is present to ensure the decisions made by actors participating in and designing the trials are within appropriate ethical bounds (Davenport & Bean, 2023). Giving AI the role of ensuring moral compliance generates a positive task–capability fit effect, reducing the aversion towards AI's involvement in morally relevant situations because consumers perceive AI to have reliable and consistent performance (Castelo et al., 2019; Everett et al., 2016).

As we argued above, the assumption is also often made that there is something uniquely human about moral decisions (exemplified by moral dilemmas above) and therefore without a human mind, moral decision-making is impossible to conceptualize (Gray et al., 2012). Much of this argument, however, is optimized for moral decision-*making*. However, when the research focus is on moral compliance, the mere upholding of moral decisions already made by someone else, one may no longer need to assume that a human mind is the only appropriate compliance agent. Indeed, recent research shows that moral compliance may occur semi-automatically, without assuming a mind (Gächter et al., 2025).

Consequently, AI may fit the task domain of moral compliance effectively, resulting in appropriate outcomes (Castelo et al., 2019). There is no need to assume that an agent tasked with moral compliance requires a "uniquely human" character (Haslam, 2006), which a full-scale moral decision-making process is argued to presuppose. Indeed, one may be perceived as an appropriate moral agent in this sense (for moral compliance tasks), without needing to have a humanlike mind (Gray et al., 2012).

While it may not need a uniquely human mind, moral compliance can still be complex, with a large number of rules, data sources and factors to consider. For instance, moral compliance involving the allocation scarce resources comes with comprehensive regulations and principles to ensure that a perception of fairness is generated among the agents, stakeholders, and observers (Bartels et al., 2015). Especially in business ethics contexts, the complexity of regulations, the number of variables, and the availability of information increase the perceived cognitive burden (Campbell & Winterich, 2018). AI's perceived information-processing superiority over humans suggests that, in moral compliance settings, not only would consumers not reject AI, but quite the opposite, they may prefer them over humans.

Consumers' firsthand experience with AI-powered products and services may strengthen the perception of AI's fit for moral compliance (Said et al., 2023). Virtual personal assistants (e.g., Siri) operate based on receiving users' commands, such as for playing music or scheduling appointments. Their responses are also often constrained by predefined rules, which are designed to comply with universal moral principles as well as legal and regulatory requirements and orders. A similar observation can be made of generative AI (such as ChatGPT) and its ethical guardrails. These AIs can refuse consumers' prompts should they deem them to be in violation of predefined rules. Therefore, consumers have already been exposed to AIs in moral compliance roles.

Businesses have also started using AI in corporate moral compliance task settings. Indeed, AI has been assisting with regulatory compliance designed to protect stakeholders and shareholders (Munoko et al., 2020; Xu et al., 2023). Companies have also sought AI's assistance in corporate governance – for example, in the domain of Environmental, Social, Governance (ESG) metrics (Ehret, 2023). These examples highlight how AI is emerging as a valuable tool for ensuring ethical accountability and promoting transparency in corporate practices, which, when observed by consumers, should result in higher consumer evaluations compared with humans. Taken together, we expect:

> **H1:** When AI has a role of upholding moral compliance, it generates more positive consumer evaluations compared with when humans have such a role.

**The Mediating Role of Motivational Inferences**

The explanation for our proposed effect is deeper than mere fit. As social beings, humans are naturally motivated to understand each other's behaviors in depth because this information tends to be consequential for their lives (Kelley & Michela, 1980). As part of making sense of the social world and others' behaviors, humans infer intentions, or objectives that a counterparty may have in mind before taking action (Reeder, 2009). The attribution of intentions occurs toward both living and non-living entities in any social setting, including marketplace exchanges. For example, when receiving an offer from a marketer that is better or worse than initial expectations, consumers infer the intentions of the administering agent, which subsequently influences consumers' decisions on the offer (Garvey et al., 2023). This notion of attribution of intent even to non-living things fits into humans' well-documented, general anthropomorphizing tendencies (Epley et al., 2007; Yang et al., 2020).

Similarly, in the moral domain, individuals regularly infer others' motives (Bucciarelli et al., 2008; Malle, 2021). Motives inform observers of the underlying reason for an actor's particular actions (Carlson et al., 2022). Inferring motives helps with the moral calculus of forming judgments about the causes of an actor's behavior, their intentions, and ultimately, their moral character (Bigman & Tamir, 2016). Through inferring motives, people evaluate whether an actor's behaviors were driven by a desire to promote fairness, a genuine concern for others' well-being, or, quite the opposite, self-interest (Miller, 1999; Tyler, 1990; Ratner & Miller, 2001). These inferred motives are subsequently connected to the moral character of the actor (Reeder 2009). Inferred motives also inform perceptions of both an action's righteousness and an actor's trustworthiness (Carlson et al., 2022). For instance, violent acts may be seen as more permissible if observers perceive self-defense as the motive rather than malicious intent (a type of moral inference-making enshrined in legal frameworks in many jurisdictions around the world). Even in the absence of observable outcomes, inferring motives facilitates the assessment of moral character (Hartman et al., 2022; Walker et al., 2021). For instance, people may judge someone's moral character negatively if they infer the person possesses a desire to harm others, even if no actual harm was committed (Cushman, 2008).

Inferring morally relevant motives can occur toward non-human entities (Malle, 2016). Given its increasing presence in everyday life, humans likely infer the motives behind an AI's actions and decisions (Pataranutaporn et al., 2023). Popular media (such as the movies *2001: A Space Odyssey; The Matrix; The Terminator, Transcendence; etc.)* have popularized the notion that AI may, under some circumstances, act according to ulterior motives, such as financial interests or survival motives, and may have "intentions" to cheat (Puntoni et al., 2019; Szollosy, 2017).

However, while humans can clearly entertain the thought of ulterior AI motives, such anthropomorphizing inferences are balanced by the obvious fact (and matching inferences) that AI is a non-living entity (Pataranutaporn et al., 2023). This biological distinction between humans and AI means that AI self-evidently does not hold *desires* or *needs* for wealth, status, or power that commonly imbue humans with ulterior motives (Perry, 2023). Being a non-living entity also means AI cannot have, and will not reap the benefits of exercising, ulterior motives, such as avoiding responsibilities or committing fraud for financial gain (Greene, 2014). In compliance settings, in particular, where moral rule application is clearcut and lacks ambiguity, human observers will more likely rely on the belief that AI does not and cannot have ulterior motives than the opposite. In line with this argument, human observers indeed recognize AI's inherent impartiality (Claudy et al., 2022).

In contrast, over millennia, *humans* have a mixed record of upholding moral standards and norms (Harari, 2014). A cynic may argue that the very need for upholding such rules signals that humans cannot be trusted morally. Indeed, the mere knowing of what is right and wrong does not guarantee that humans consistently act ethically, as a range of factors can derail humans from doing the right thing, even if they know the correct moral course of action (Monin & Merritt, 2012). In the domain of business ethics, Volkswagen's Dieselgate scandal highlights human greed and peer pressure that paved the way for deviant behaviors to cheat emission guidelines and standards (Hotten, 2015). Business corruption, enacted exclusively by humans, remains a persistent problem, exacerbated by strategic behavior to cover up ulterior motives (Bullough, 2018; Kouchaki, 2011). In sum, humans are widely perceived as weak in moral compliance in the face of self-interest and societal pressures. Thus, our second hypothesis is:

> **H2**: The positive main effect of AI (vs. humans) moral compliance roles on consumer evaluations is mediated by lower inferences of ulterior motives for AI (vs. humans).

**Overview of Studies**

We tested our hypotheses with five experiments. The first study provides evidence for the main effect of deploying AI versus humans in moral compliance roles on consumer evaluations, finding a positive effect in support of H1. For further generalizability, Study 2 replicates the main effect observed in Study 1 using a different sample, two new types of product category (with high vs low moral risk salience), and a new type of incentive for moral lapsing (tactical rather than strategic) for robustness. The mediating role of motivational inferences is demonstrated through indirect effects in Study 3 (varying the DV to include both attitudinal and behavioral intent outcomes) and process-by-moderation in Study 4. Finally, Study 5 directly contrasts human inferences about AI's role in moral compliance vs moral decision-making.

In the design of our stimulus materials for all studies, we ensure that we frame the compliance task in a context where ulterior motives could be seen to play a role. This context is critical to our research question, as we are not focused on whether AI is simply better at following rule-based decision-making, which would be relevant to any decision context, but rather that it is better at following rule-based decision-making in compliance contexts, where compliance codes are put in place specifically to guard against individuals or organizations otherwise operating out of self-interest. To this end, we first set the "moral scene," making it clear there is a tension between adhering to compliance standards (moral decision) vs. the performance or profitability of the organization (self-interest decision).

Second, we show scenarios where the consequences of intentional moral action are clear and where observer inferences about these intentions (including ulterior motives) can play out. We focus on the possibility for inferences for intentions because such perceptions have been shown to form the core of moral inference-making (Alicke, 2000; Cushman & Greene, 2012; Malle, 2021; Malle et al., 2014; Malle & Robbins, 2025).

Third, we pay close attention to avoid demand biases. After the setting of the moral scene, we place the "moral actors" on the stage and ensure that we never make any claims about their behavior – we leave that assessment entirely up to the participant observers (like for an audience in a theatre). Relatedly, it is important to note, we never judge or make claims of actual human behavior on stage (whether humans objectively would act less morally than AI) – all our findings relate to audience inferences.

Fourth, we equalize the AI vs human (between subject) conditions (without sacrificing inherent differences) in five ways: (1) giving both agents a singular identity (e.g., "Evalu8" vs. one Executive, rather than vague "algorithms" or "the committee"); (2) balancing anthropomorphism by naming the AI but not the human – a conservative choice, as it increases humanlike attributions to the AI; (3) aligning perceived agency by emphasizing AI "self-learning" to parallel humans' natural autonomy; (4) holding competence constant by stating that both agents receive the same training; and (5) equating functional, task-relevant performance via an identical detailed description of what the agent actually did, varying only whether it is labelled AI or human.

Fifth, we measure, immediately after the scenarios, the consequential outcomes (in our case, consumer purchase intentions from the focal business) and subsequently, evidence about the mechanism (in S3 and S4). Finally, we collect controls and demographics. Web Appendices A-E contain stimuli and measures used in all studies, as well as additional analyses.

**Study 1: Main Effect**

Study 1 tested the hypothesis that deploying an AI (vs. a human) as a moral compliance agent increases consumers' evaluations (H1). This study used a consumer products context (athletic shoes) in which participants read about the involvement of either an AI or human agent in assuring compliance in a hypothetical company's ethical sourcing and manufacturing processes. Consumer evaluations were operationalized as purchase intentions.

*Participants and Design*

A total of 300 US residents (US, UK, Ireland, Australia, Canada, and New Zealand nationalities, all English native speakers) were recruited from Prolific, an online research panel platform, to participate in the study. The participants were then randomly assigned to one of the two (AI agent vs. human agent) between-subjects conditions.

*Procedure*

Participants in both conditions were first asked to read a short article on a contemporary issue in the corporate world involving Environmental, Social, and Governance (ESG) practices to set the moral scene (identical in both conditions). We selected ESG practices among corporations as context given the importance of the issue for consumers, businesses and societies more broadly (Noble, 2023). Respondents, then, encountered the stimuli depicting Company XYZ, an athletic shoe-producing company, which has taken steps to ensure its sourcing and manufacturing processes to be as ethical as possible. In one condition the focal agent to ensure moral compliance was an AI (Evalu8, a self-learning AI system); in the other, one of the firm's human executives. In both conditions, participants were told that the focal agent had completed the necessary ethical training. Further, in both conditions, identical descriptions detailed the functional tasks this compliance meant (checking aspects of ethical manufacturing in various geographic locations where the company operates, which was labelled a complex endeavor). In the end, again using identical text, the condition highlighted the moral hazard inherent in the situation, the presence of financial incentives. The only difference between the two conditions was if an AI or an executive had the moral compliance agent role.

After reading the stimuli, participants indicated their purchase intentions on the athletic shoes produced by the company on three items ($\alpha = .96$), all on seven-point Likert scales. As a control variable, we also measured participants' pre-existing perceptions of ethical consumption with two items on seven-point Likert scales, as past studies showed this variable influenced individuals' purchase evaluations ($\alpha = .96$; Paharia, 2020). At the end of the study, participants responded to demographic questions (such as gender, age and political philosophy) and a manipulation check.

*Results and Discussion*

To control for attention, from the 300 recruited participants, we omitted those who spent less than 10 seconds reading the short article at the beginning of the study (and in all subsequent studies on Qualtrics we forced a minimum 10-second exposure). This resulted in 259 participants ($M_{age}$=39.1 years; $SD_{age}$=14.7 years; 51% female). A chi-square test revealed a significant difference between the two agent types $\chi^2(1, N = 259) = 243.26, p < .001$), suggesting the manipulation of agent type worked.

A t-test showed a significant effect of agent type on participants' intentions to purchase the athletic shoes ($t(257) = -2.09, p = .037$, Cohen's $d = 0.26$). Specifically, participants in the AI condition ($M_{AI}$ = 4.88, $SD_{AI}$ = 1.36) indicated higher intentions to purchase the athletic shoes than those in the human condition ($M_{human}$ = 4.53, $SD_{human}$ = 1.28). Figure 1 depicts these findings.

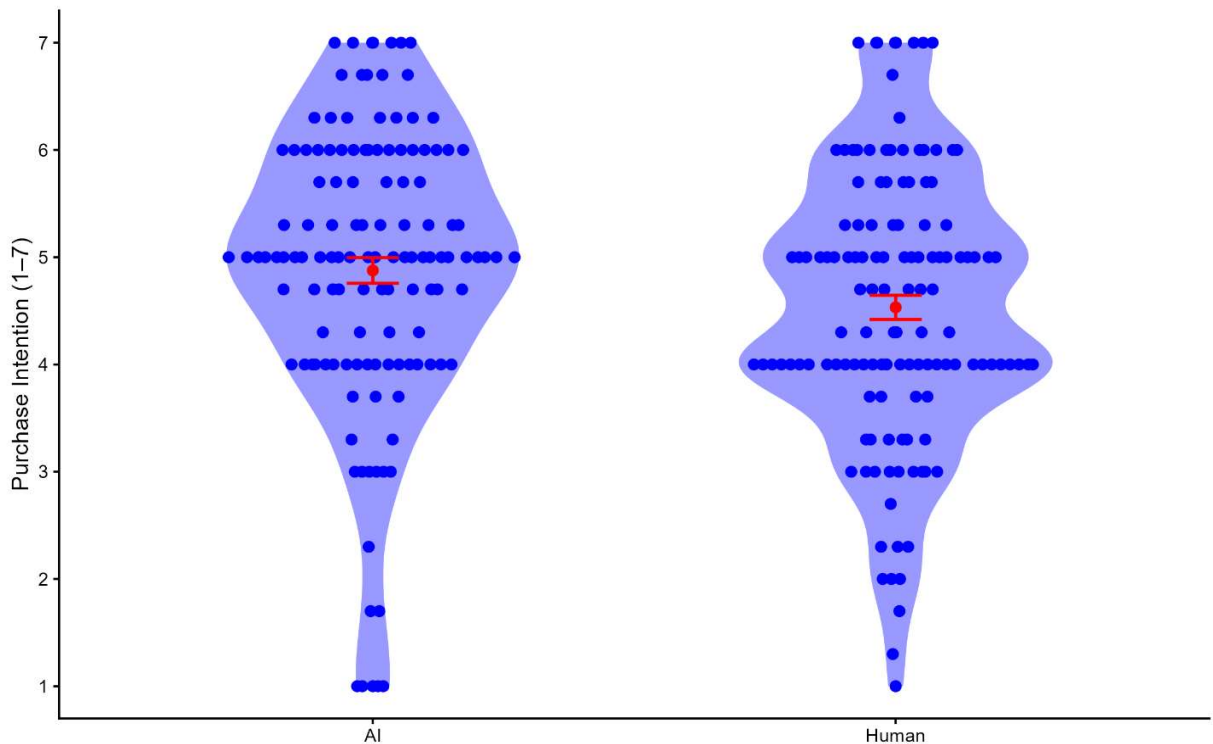

**Fig. 1** The effect of agent type on purchase intentions. *Note:* Violin plots with beeswarm dots for individual cases, showing the full distribution, produced with the ggplot and ggbeeswarm packages in R. The contours of the violins represent the kernel density estimate of the data's distribution for each group at every value on the y-axis (a wider contour means more data are clustered around that y-value, a narrower section means fewer data points there). Each beeswarm dot represents an individual observation in the data. The red dots indicate means and the whiskers around them are ±1 Standard Errors.

To control for pre-existing attitudes towards the moral issue at hand, we included perceptions of ethical consumption as a covariate. The ANCOVA test indicated that our main predictor, agent type remained significant after controlling for this variable ($F(1, 256) = 4.49$, $p = .035$). This covariate was also a significant predictor ($F(1, 256) = 7.11$, $p = .008$).

Study 1 shows that using an AI (vs. a human) agent in a moral compliance role in ethical manufacturing processes increases participants' purchase intentions. This finding provides initial evidence supporting H1. Notably, the main effect of agent type on purchase intentions remains significant when we control for participants' perceptions of ethical consumption.

**Study 2: Replication of the Main Effect and Robustness**

Building on Study 1, Study 2 aims to replicate the effect and increase its generalizability. First, in some product categories ethical risks are quite top of mind. For example, ethical sourcing issues in clothing manufacturing are both well-publicized and salient to consumers (McKinsey, 2024), while in other categories (e.g., consumer staples such as milk) moral risks are possible but not salient. To assure that it is not only product categories where moral risk is salient in which our effect holds, we wanted to see if our effect replicates in both higher (T-shirts) vs. lower (milk) moral risk-salient product categories. Second, we used a different sample (university students), again, for broader generalizability and robustness. Third, we also changed the type of incentive firm actors were presented to have for lapsing on moral compliance, shifting from strategic gains in Study 1 (improving overall profitability), to tactical/operational ones (preventing temporary operational disruptions). If our effect holds even for the context of tactical gains, it proves our study's broader applicability.

*Participants and Design*

A total of 396 undergraduate students from an Australian university participated in this study, recruited via a student research pool. The participants were then randomly assigned to one of four conditions in a 2 (Agent: AI vs. human) × 2 (Product category: Low moral risk saliency (milk) vs. High moral risk saliency (T-shirt) goods) between-subjects design.

*Procedure*

Participants began by reading a sensitization article similar to Study 1 for at least 10 seconds. Then, in the low (vs high) risk-salient product category, they read a milk (vs. T-shirt) company's effort to make its labor practices in production as ethical as possible. Then, similar to Study 1, in both the AI and the human agent conditions equally, they were told that the agent-in-charge attended ethical training and the executives' bonuses depended on the firm's profit. In addition, in both conditions, participants were told that the company stops its operations when a decision is flagged due to moral compliance issues, which may cause negative financial implications for the company. After reading the stimuli, the participants then indicated their purchase intentions on three items ($\alpha$ = .86), all on seven-point scales (i.e., 1 = "not at all," and 7 = "very much").

As a covariate, we also measured participants' pre-existing attitudes towards AI as past studies showed it influenced individuals' evaluations (Schepman & Rodway, 2023). At the end of the study, participants responded to a manipulation check question on our main predictor, agent type.

*Results and Discussion*

Of the 396 recruited participants, 203 (193) participants were in the low (high) moral risk salient product category condition. A chi-square test revealed a significant difference between the two agent types $\chi^2(1, N = 396) = 107.33, p < .001$), suggesting the manipulation of agent type worked.

A two-way analysis of variance on purchase intentions showed a significant effect of agent ($F(1, 392) = 13.74, p < .001$), a marginally significant effect of product category ($F(1,$

392) = 3.58, $p = .059$), and no significant interaction of agent and product category ($F(1, 392) = 0.24$, $p = .63$).

Planned contrasts revealed that in the low moral risk salience category, participants in the AI condition had a higher level of purchase intentions ($M_{AI} = 5.14$, $SD_{AI} = 1.14$) than those in the human condition ($M_{human} = 4.64$, $SD_{human} = 1.26$); $F(1, 392) = 8.92$, $p = .003$, Cohen's $d = 0.42$). Similarly, in the high moral risk salience category, those in the AI condition had a higher level of purchase intentions than those in the human condition ($M_{AI} = 4.86$ vs. $M_{human} = 4.48$, $SD_{AI} = 1.21$ vs. $SD_{human} = 1.12$; $F(1, 392) = 4.98$, $p = .026$, Cohen's $d = 0.32$).

For robustness, we entered pre-existing attitudes towards AI as a covariate. The ANCOVA test showed that our predictor, agent type remained significant, after controlling for attitudes toward AI ($F(1, 391) = 13.78$, $p < .001$). The covariate was not a significant predictor ($F(1, 391) = 2.21$, $p = .14$).

Using an AI (vs. a human) agent in moral compliance increases observing consumers' purchase intentions across two product categories, a finding we replicate with a different population from Study 1. The results are also robust as participants' pre-existing attitudes about AI do not change the main effect. Further, in this study, we changed the type of gain (from strategic to tactical) that firm actors may obtain from lapsing in moral compliance and found the same effect as in Study 1. These findings suggest that the effect generalizes across various populations, various moral risk-level product categories and various moral compliance lapsing scenarios.

**Study 3: Mediation**

The objective of Study 3 was to provide initial process evidence through demonstrating indirect effects. As our theorization posits, the mechanism through which AI's edge over humans materializes in moral compliance is the inference of inherently lower ulterior motives for AI. A secondary objective was to further increase generalizability by (1) changing the company's consumer offerings from products to services (consumer banking), (2) changing the operationalization of the consumer evaluations DV from a purely behavioral intention measure to a composite of behavioral (purchase intentions) and attitudinal (perceptions of appropriateness) index, (3) adding new individual differences controls such as familiarity with the specific service category in question as well as generalized technology affinity.

*Participants and Design*

A total of 299 US residents (US, UK, Ireland, Australia, Canada, and New Zealand nationalities, all English native speakers) were recruited from Prolific. The participants were then randomly assigned to one of the two (AI vs. human moral compliance agent) between-subjects conditions.

*Procedure*

Similar to our previous studies, participants began by reading a sensitization article related to ESG practices among companies. Then, they encountered similar stimuli as used in Studies 1 and 2, except for a difference in industry context (consumer financial services instead of products). In both conditions, participants were told that the company compensates its executives based on profits achieved. In this study, we operationalized the DV, consumer evaluations, with both attitudinal (whether consumers perceived the agent's moral compliance actions appropriate, three items adapted from Bigman & Gray, 2018) and behavioral measures (intentions to bank with the institution, three items similar to Study 1 and 2). These items were highly reliable as a scale ($\alpha$ = .92). Agents' perceived ulterior motives served as the mediating variable (seven-point Likert scale, three items, $\alpha$ = .82; adapted from Ferraro et al., 2013). Finally, as controls, we also measured participants' familiarity with the banking industry context on a single item and their generalized technology affinity on two items ($\alpha$ = .93; both adapted from Castelo et al., 2023). At the end of the study, participants responded to the same demographic questions and manipulation check item as in Study 1.

*Results and Discussion*

A chi-square test of 299 responses ($M_{age}$=44.2 years; $SD_{age}$=13.6 years; 50% female) reveals a significant difference between the two agent types $\chi^2$(1, N = 299) = 256.59, *p* < .001), suggesting the manipulation of agent type worked as before.

To test the main effect, we conducted a t-test using agent type as the factor and consumer evaluations as the dependent variable. The test revealed a significant main effect (*t*(297) = – 6.29, *p* < .001, Cohen's *d* = 0.73). Specifically, participants in the AI condition ($M_{AI}$ = 5.24, $SD_{AI}$ = 1.35) indicate a higher evaluation than those in the human condition ($M_{human}$ = 4.21, $SD_{human}$ = 1.47).

We started testing the proposed mediation hypothesis by analyzing the main effect of agent type on the mediator, perceived ulterior motives. Confirming expectations, consumers inferred that the AI agent had lower ulterior motives ($M_{AI}$ = 3.02, $SD_{AI}$ = 1.60) than the human agent ($M_{human}$ = 4.36, $SD_{human}$ = 1.39; *t*(297) = 7.71, *p* < .001, Cohen's *d* = 0.89). For indirect effects, we ran PROCESS Model 4 with 5,000 bootstrapping resamples and a fixed seed value for reproducibility (Version 4.3.1 for R; Hayes, 2022). Results reveal a significant indirect effect of agent type on consumer evaluations mediated by ulterior motive inferences (*b* = 0.28, BootSE

= 0.05, with a 95% bootstrap confidence interval that did not include zero [0.18, 0.38], full model in Figure 2 and Web Appendix C).

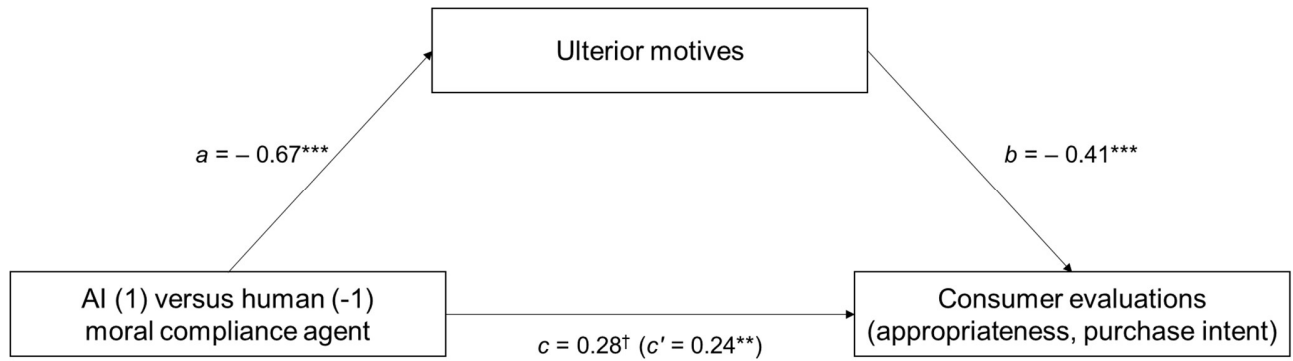

**Fig. 2** Ulterior motives mediate the effect of moral compliance agent type (ai vs. Human) on consumer evaluations. *Note*: *p < .05, **p < .01, ***p < .001; †significant at 5% level based on 95% bootstrapped confidence intervals with 5,000 repetitions; coefficients *a*, *b*, *c*, *c'* are unstandardized; indirect effect is *c*; direct effect is *c'* and is in parentheses

The observed indirect effect is robust after controlling for familiarity with banking services and generalized technology affinity or both. Similarly, the effect persists when testing the behavioral and attitudinal components of our consumer evaluations index separately. Details of these robustness tests are available in the Web Appendix C.

In Study 3, we replicated findings from our previous two studies in a different domain (i.e., consumer services) in which using an AI (vs. a human) moral compliance agent increases consumer evaluations. The results of Study 3 also support our hypothesis H2, demonstrating that ulterior motive inferences serve as the underlying mechanism driving the main effect. Our effect is robust after operationalizing consumer evaluations as attitudinal or behavioral and after controlling for individual differences (familiarity with industry and general technology affinity).

**Study 4: Process-by-moderation**

In Study 4, we used a process-by-moderation approach (Spencer et al., 2005), which is a method to provide additional evidence about the mechanism we theorize (ulterior motives), beyond the indirect effects demonstrated in Study 3. Accordingly, we directly manipulated consumers' inferences about the moral compliance agents' ulterior motives to provide further support for the proposed theoretical mechanism. If the effect of agent type on consumer evaluations is indeed driven by lower inferred ulterior motives in the human condition than the AI condition, we expect that the main effect will be blocked when we set these inferences to be high, by directly informing respondents about unequivocal ulterior motives for each agent in their respective conditions.

*Participants and Design*

A total of 601 US residents (US, UK, Ireland, Australia, Canada, and New Zealand nationalities, all English native speakers) were recruited from Prolific. The participants were then randomly assigned to one of four conditions in a 2 (Agent: AI vs. human) × 2 (Ulterior motives inferred: High vs. control) between-subjects design.

*Procedure*

Participants began by reading an article similar to our previous studies. Then, similar to Study 1, they encountered the stimuli in which either AI or human moral compliance agents are deployed to oversee the sourcing and manufacturing processes in producing athletic shoes. In the high ulterior motive condition, we manipulated participants' inferences by telling them that the moral compliance agent (i.e., either AI or human) appointed by the company was previously found prioritizing financial gains at the expense of ethical guidelines (i.e., high ulterior motive condition). In the control condition, no such information was provided about the moral compliance agent. Similar to our previous studies, participants were told that the company compensates its executives based on the firm's financial performance in both conditions equally.

After reading the stimuli, the participants indicated their purchase intentions on three items ($\alpha$ = .97), all on seven-point scales (i.e., 1 = "not at all," and 7 = "very much"). Then, they responded to demographic question items and two manipulation check items checking on each of the manipulations in the 2 × 2 design.

*Results and Discussion*

We ran analyses with all responses ($M_{age}$=44.1 years; $SD_{age}$=15.1 years; 49% female). To test if the agent type manipulation worked, we ran a logistic regression with the agent type and ulterior motivation factors as predictors, including their interaction, on the manipulation check variable for agent type. The main effect of agent type was significant and in the expected direction ($b$ = 7.24, SE = 0.74, $z$ = 9.79, $p$ < .001), indicating that participants correctly classified the agent type. Neither the main effect of ulterior motivations ($b$ = − 0.52, SE = 0.74, $z$ = − 0.71, $p$ = .478), nor the interaction term ($b$ = 0.57, SE = 1.11, $z$ = 0.52, $p$ = .605) was significant, confirming that the agent type manipulation worked as planned.

To test if the ulterior motives manipulation worked as planned, we ran a 2 × 2 ANOVA, with agent type and manipulated ulterior motives and their interaction as factors, and the scaled manipulation check variable for ulterior motives as DV. Results showed that the manipulation had the intended main effect ($F(1, 597)$ = 144.21, $p$ < .001) and the expected agent type main effect on the manipulation check item measuring ulterior motives ($F(1, 597)$ = 42.46, $p$ < .001). The interaction term, however, was not significant ($F(1, 597)$ = 0.91, $p$ = .34). Planned contrasts revealed that respondents inferred lower ulterior motives for AI versus humans in both the high manipulated ulterior motives condition ($M_{AI\ x\ HighUM}$ = 4.84, $SD_{AI\ x\ HighUM}$ = 1.76 vs. $M_{Human\ x}$

$_{HighUM}$ = 5.34, SD$_{Human\ x\ HighUM}$ = 1.37; $F(597)$ = 15.99, $p$ <.001, Cohen's $d$ = – 0.46) and in the control (M$_{AI\ x\ ControlUM}$ = 2.89, SD$_{AI\ x\ ControlUM}$ = 1.56 vs. M$_{Human\ x\ ControlUM}$ = 3.64, SD$_{Human\ x\ ControlUM}$ = 1.72; $F(597)$ = 29.24, $p$ <.001, Cohen's $d$ = – 0.67).

Next, we tested the main interactive hypothesis of interest, whether a high ulterior motivation manipulation removes the AI vs human moral compliance agent differential on consumer evaluations (purchase intentions). A 2 × 2 ANOVA revealed that both the agent type ($F(1, 597)$ = 6.16, $p$ = .013), and the manipulated ulterior motivation factor had a main effect ($F(1, 597)$ = 130.20, $p$ < .001). However, unexpectedly, the interaction was non-significant ($F(1, 597)$ = 0.003, $p$ = .957). Planned contrasts confirm that respondents infer marginally lower ulterior motives for AI versus humans in both the high manipulated ulterior motives condition (M$_{AI\ x\ HighUM}$ = 3.49, SD$_{AI\ x\ HighUM}$ = 1.55 vs. M$_{Human\ x\ HighUM}$ = 3.19, SD$_{Human\ x\ HighUM}$ = 1.59; $F(597)$ = 3.52, $p$ = .061, Cohen's $d$ = 0.22) and in the control (M$_{AI\ x\ ControlUM}$ = 4.85, SD$_{AI\ x\ ControlUM}$ = 1.23 vs. M$_{Human\ x\ ControlUM}$ = 4.53, SD$_{Human\ x\ ControlUM}$ = 1.38; $F(597)$ = 3.33, $p$ = .069, Cohen's $d$ = 0.21). Figure 2 depicts this pattern of results.

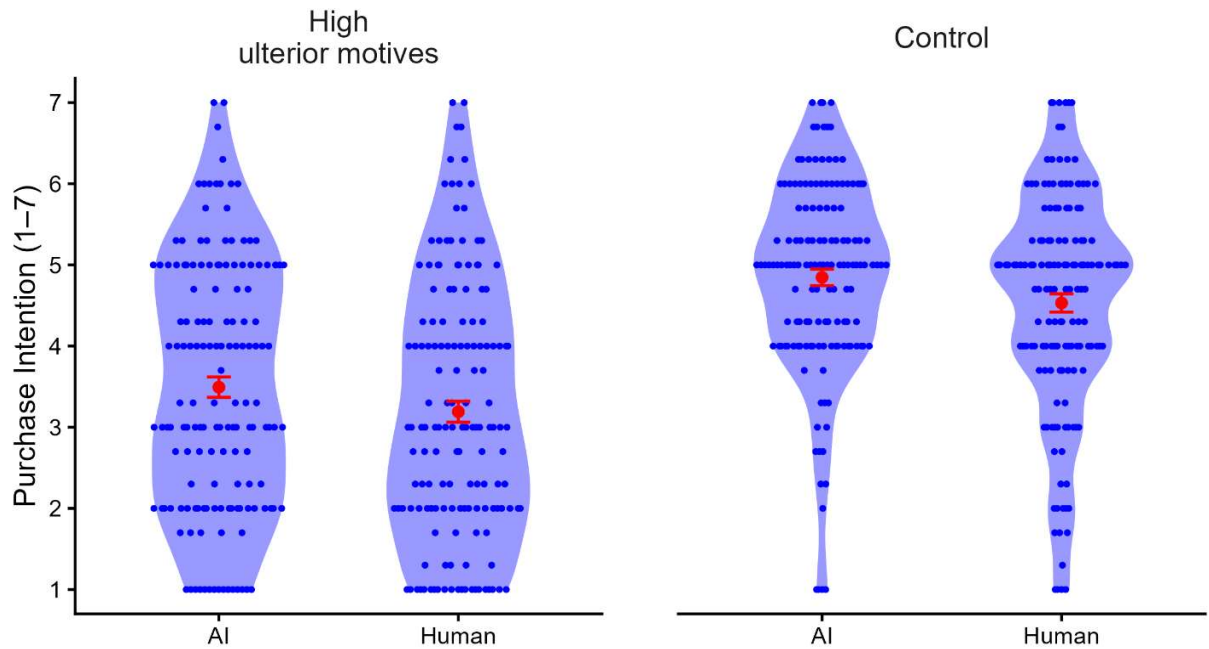

**Fig. 2** The interactive effect of agent type and forced ulterior motive perceptions on purchase intentions. *Note:* Violin plots with beeswarm dots for individual cases, showing the full distribution, produced with the ggplot and ggbeeswarm packages in R. The contours of the violins represent the kernel density estimate of the data's distribution for each group at every value on the y-axis (a wider contour means more data are clustered around that y-value, a narrower section means fewer data points there). Each beeswarm dot represents an individual observation in the data. The red dots indicate means and the whiskers around them are ±1 Standard Errors.

Given that we failed to reject the interactive null hypothesis but found a similar pattern of main effects of AI versus human agent type on ulterior motives and purchase intentions, we wanted to test, in an exploratory fashion, whether we can replicate the indirect effect from Study 3, even after controlling for our ulterior motives intervention. We ran PROCESS Model 4 with 5,000 bootstrapping resamples and a fixed seed value for reproducibility (Version 4.3.1 for R; Hayes, 2022). Results revealed a significant indirect effect of agent type on consumer evaluations, mediated by the ulterior motivations manipulation check ($b = 0.20$, BootSE = 0.04, with a 95% bootstrap confidence interval that did not include zero [0.13, 0.27]). This effect persisted even after controlling for the manipulation of ulterior motives ($b = 0.16$, BootSE = 0.03, with a 95% bootstrap confidence interval that did not include zero [0.11, 0.22]).

We attempted to directly manipulate participants' inferences on moral compliance agents' ulterior motives in this study, in line with the process-by-moderation logic (Spencer et al., 2005). Although the manipulation to set ulterior motives artificially high did influence respondents' ulterior motive inferences, it was unable to overcome what appears to be a deep-seated belief that humans are more prone to ulterior motives: Respondents scored ulterior motives higher for humans than AI, irrespective of our ulterior motives-equalizing manipulation.

While this is technically a problem for Spencer-style moderation-for-mediation analysis, it does show that people are quite ready to activate the human ulterior motives/self-interest norm (Ferraro et al., 2013; Ratner & Miller, 2001), which we believe is what is behind the efficacy of allowing AI to act in the role of moral compliance. Further, when incorporating our blocking manipulation as a covariate, we again replicated our indirect effect finding from Study 3, which suggests that such ulterior motivation inferences mediate the effect of AI vs. human moral compliance on purchase intentions.

Across Studies 1 to 4, we show that positioning an AI, rather than a human, as the moral compliance agent reliably elevates consumer evaluations across product and service settings, different moral-risk contexts, incentive structures, and outcome measures, and that this advantage is systematically underpinned by lower inferences of ulterior motives for AI agents. Yet, these studies all examine cases where the agent's role is to uphold pre-existing rules, leaving open whether AI's apparent moral advantage is unique to moral compliance or reflects a more general preference for AI over humans, even when AI is asked to decide what is morally acceptable (moral decision-making). To address this remaining gap, in Study 5 we directly contrast reactions to AI versus human agents placed either in a moral compliance role or in a moral decision-making role, allowing us to test whether consumer acceptance of AI is indeed contingent on the humble, rule-enforcing role we propose.

**Study 5: Comparing moral compliance and decision-making**

The main objective of Study 5 was to directly compare moral compliance and decision-making. In addition, we wanted to rule out a possible confound that the word "executive," used at the end of prior scenarios, may have unduly influenced the results. Instead, we presented the moral dilemma as that of the "company." Further, we removed the sensitization article and instead provided a shorter description of the moral dilemma agents could face, which as noted above, is essential to our study design in order to provide a test our research question. In the scenarios themselves, we provided more detailed information about what each term, moral compliance and moral decision-making, means concretely. Finally, we also added a human name ("Everett") to the human agent, to rule out any possible confounds arising from the differences in prior scenarios where AI had a name and the human executive did not (stimulus materials, measures are included in Web Appendix E). The study was pre-registered at aspredicted.org (https://aspredicted.org/g67un4.pdf)

*Participants and Design*

A total of 302 US residents were recruited from Prolific. The participants were randomly assigned to one of two conditions (Moral compliance vs. Moral decision-making) in a between-subjects design.

*Procedure*

Participants started by reading an introductory section, asking them to imagine a scenario where a large global clothing company that sells T-shirts and other garments wanted to improve its ethical behavior by paying everyone fairly and avoiding forced or "slave" labor in its supply chain (identical in both conditions). Briefly, the moral dilemma was presented: There may be financial incentives to cheat, because relaxing labor practices (e.g., allowing questionable suppliers) would make it easier and cheaper to produce clothes and so would increase profits; but this would also result in the company acting unfairly and unethically. Conversely, being very strict about labor practices (e.g., cutting ties with suppliers that raise concerns) would make it more difficult and expensive to produce clothes and so would reduce profits, but the company would be acting more fairly and ethically.

Next, respondents were exposed to one of two conditions. In the moral compliance condition, they were informed that the company had already committed to external moral rules and standards for labor practices that clearly spell out what counts as acceptable and unacceptable treatment of workers. The focal moral compliance role was to ensure adherence to moral rules and carry out routinized applications in a consistent manner. In the moral decision-making condition, the company had not committed to external moral rules and standards for labor practices yet – instead, they were creating a role to define these practices using moral decision-making.

Then, as the dependent variable, a slider measure captured respondent perceptions whether in this role an AI (Evalu8) or human (Everett) would lead to more ethical outcomes. The sliders were randomly counterbalanced in direction (human to the right vs. human to the left). After transformation, the values range from 0 to 100, where 100 represents a higher preference for AI. We also collected open-ended responses, asking why respondents chose the answer they gave. Finally, we collected age, gender and prior attitudes to AI (Schepman & Rodway, 2023) as covariates.

*Results and Discussion*

Of the 302 participants, 153 were in the moral compliance and 149 in the moral decision-making condition. To compare respondent perceptions of ethical fit in the two conditions, we conducted a t-test, which yielded a significant effect ($t(300) = -2.95$, $p = .003$, Cohen's $d = 0.34$). Specifically, participants in the moral compliance condition ($M_{MC} = 46.89$, $SD_{MC} = 38.38$) indicated a higher preference for AI than did those in the moral decision-making condition ($M_{MDM} = 34.47$, $SD_{MDM} = 34.85$). Figure 3 depicts these findings.

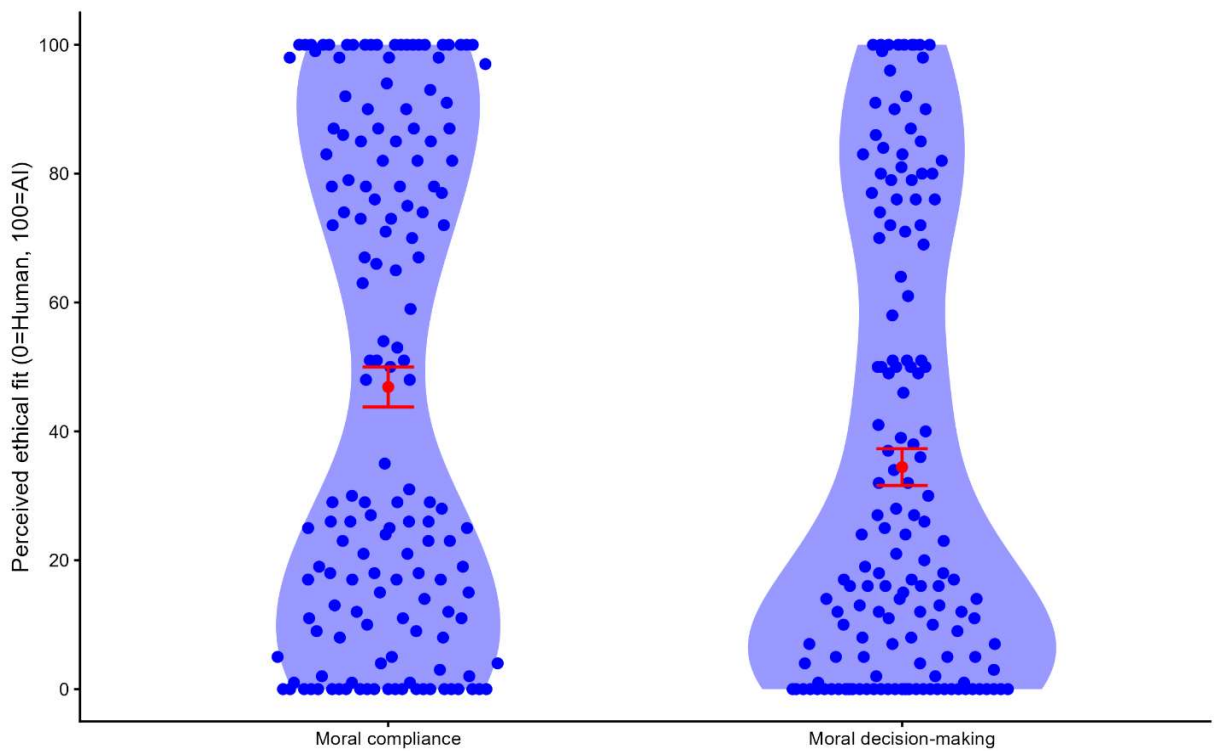

**Fig. 3** The effect of moral compliance vs moral decision-making on perceived ethicality. *Note:* Violin plots with beeswarm dots for individual cases, showing the full distribution, produced with the ggplot and ggbeeswarm packages in R. The contours of the violins represent the kernel density estimate of the data's distribution for each group at every value on the y-axis (a wider contour means more data are clustered around that y-value, a narrower section means fewer data points there). Each beeswarm dot represents an individual observation in the data. The red dots indicate means and the whiskers around them are ±1 Standard Errors.

Since pre-existing attitudes towards AI were significantly correlated with perceived ethicality in our scenarios ($r(300) = .41, p < .001$), we entered it as a covariate, with our manipulation as the main predictor and perceived ethicality as DV. The ANCOVA test showed that moral compliance vs decision-making remained significant even after controlling for prior attitudes toward AI ($F(1, 299) = 10.23, p < .002$). The covariate was a significant predictor, as well ($F(1, 299) = 62.04, p < .001$), having a positive relationship with the DV (the higher prior attitudes were, the higher perceived ethicality in the scenario was for AI). These results show that, while prior AI attitudes positively influence perceived ethicality, our focal hypothesis holds regardless.

We also investigated the exploratory open-ended question "Please tell us why?" immediately following the focal DV, perceived ethicality. We used principles of inductive textual analysis (Ericsson & Simon, 1984; Gioia et al., 2013; Strauss & Corbin, 1998), and identified two core themes in the corpus: *lack of ulterior motives* (the perceived advantage of AI systems for lacking self-serving biases) and *moral context uniqueness* (the idea that moral decisions need humans due to the emotionality, subjectivity and complexity inherent in moral situations). We created a variable with three levels: two levels capturing the above themes and an "other" category. Details of the coding procedure are in Web Appendix E.

We then compared the two conditions with regards to the prevalence of each of these themes. A chi-square test revealed a significant difference among the reasons respondents provided for their ethicality judgments across the two conditions ($\chi^2(1, N = 302) = 14.40, p < .001$). AI's lack of ulterior motives was cited significantly more in the moral compliance condition (35%) than in the moral-decision-making condition (18%). Conversely, moral context uniqueness was much more prevalent in the decision-making condition (52%), than in the compliance condition (33%).

In summary, Study 5 confirmed that human observers find AI more acceptable in moral compliance roles than in moral decision-making roles. In addition, we found more evidence for the mechanism explaining this acceptance within respondents' emic, or unprompted, lay theoretical explanations. While in decision-*making* settings, thoughts of the uniqueness of the moral contexts dominate (requiring a "human touch"), in moral *compliance* another lay theory emerges in relative importance: AI agents are deemed acceptable because, due to their perceived lack of ulterior motives, they ought to be able to better control the very human risk of loosening compliance standards out of self-interest.

## General Discussion

Over the past decade, companies across many industries, from manufacturing to healthcare, have started to integrate AI into their business activities. The proliferation of AI in society and everyday life offers both promises and caveats for businesses using it, particularly regarding how this may impact consumers. For instance, while AI improves efficiency in processing loan and job applications, companies generally do not highlight it to consumers because people remain skeptical about its decision-making abilities (McKendrick & Thurai, 2022). Algorithm aversion, especially in morally charged domains, perceived as ill-fitting for AI deployment, remains a persistent business ethics problem (Castelo et al., 2019; Dietvorst & Bartels, 2022).

However, as this present research demonstrates, moral *compliance* is a unique context – one where consumers are willing to accept the involvement of AI. Contrary to what earlier work on consumers' reactions to moral AI would predict, using AI rather than a human agent in this role *positively* influences consumer evaluations. The mechanism driving this positive effect is the attributional thought process that suggests AI agents have lower ulterior motivations compared to humans.

Five studies in both product and service contexts provide support for our theory. In each study, having an AI agent overseeing moral compliance results in higher consumer evaluations. Study 1 showed that consumers demonstrated higher purchase intentions for a pair of ethically made athletic shoes produced by a company that uses AI rather than a human agent overseeing the ethicality of its production processes. To expand generalizability, in Study 2, we examined consumer responses to the use of AI (vs. human) moral compliance agents in different product categories (i.e., low vs high moral risk saliency), moral hazard types (strategic vs tactical gain) and a different sample set (i.e., university students). Study 3 revealed that the inference of ulterior motives mediates the effect. Participants perceived that AI moral compliance agents had lower ulterior motives, which in turn led to higher consumer evaluations than in the case of humans. This effect was robust across attitudinal/perceptive (appropriateness) and behavioral (purchase intention) outcomes. In Study 4, supplied further evidence about the mechanism showing higher ulterior motives for humans agents than AI, even when we attempted to block this attribution by making it artificially high. It appears that consumers have a deeply seated belief that, when given the chance, human moral compliance agents fail more readily than AIs would. Such inferences indeed fully mediated the relationship between agent type and evaluative outcomes. Lastly, in Study 5 we showed that there is a difference between moral compliance and moral decision-making inferences: respondents accept AI in moral compliance more than in moral decision-making – because inferences about AI's lack of ulterior motives become more prominent.

**Theoretical Implications**

Our work offers three contributions to the current dialogue on behavioral approaches to AI ethics. First, it introduces the concept of moral compliance (Shirai & Watanabe, 2024). Past literature has primarily focused on AI as a decision-*maker* in the technology's interactions with consumers, delivering medical diagnoses (Longoni et al., 2019), administering discounts to consumers (Garvey et al., 2023) and determining insurance premiums and payouts (Dietvorst & Bartels, 2022). In all these morally charged situations, AI is positioned in the "front end" of the decision-making process; in essence, making the decision, replacing the human. In this study, in contrast, we position AI towards the end of the decision-making process as an agent upholding moral rules and norms that others (typically a human, humans or larger social groupings) have already made (Geva, 2000). By doing so we broaden the scope of what phenomena are to be considered under the "moral" label when AI is involved (moral compliance as much as moral decision-making) and rethink assumptions about how "essentially" human, therefore, moral agency really needs to be (Wilson & Haslam, 2009; Gray et al., 2012; Schmitt, 2020).

Second, our study contributes to the AI–task domain congruence literature (Castelo et al., 2019). For consumption decisions that rely on subjective experience, emotions, and sensory evaluations, consumers avoid using AI because they see the technology as lacking those abilities (Longoni & Cian, 2022). In contrast, in realms of factual data and technical expertise consumers may appreciate AI (Logg et al., 2019). Quite naturally, moral decision-making is grouped together with the AI-incompatible task domains by consumers, given morality's subjective, socially constructed and "essentially" human-like nature (Gray et al., 2012).

Importantly, we add a new dimension to AI–task domain congruence by focusing on the *role* an agent plays within a given domain. While moral decision-making and moral compliance are both in the moral domain, upholding pre-existing norms and rules (compliance) is clearly a different role from initiating original decisions, especially, when such decisions test human judgment to the limit (i.e., are moral "dilemmas"). Our theory proposes and finds that the role of moral compliance fits AI quite well. The role of complying with moral rules and norms presupposes objectivity and consistency (Bartels et al., 2015) an expectation which matches people's perception of AI being reliable in following commands (Castelo et al., 2019; Glikson & Woolley, 2020). Our study highlights the importance of "role" as a critical new dimension to AI–task domain congruence, showing that while moral decision-making is often seen as incompatible with AI, the specific role of moral compliance leverages AI's strengths in objectivity and consistency, offering a fresh perspective to the literature.

Finally, we contribute by providing a new theoretical mechanism, lack of ulterior motives, for why AI can be accepted, even preferred by consumers in this domain. While lack of a full human mind, especially, the dimension of experience (Gray et al., 2007) may be perceived as an AI-shortcoming, one clear benefit arises in a compliance setting, the lack of motivation for self-dealing. As a non-living entity, AI inherently does not possess desires that commonly lead humans to harbor ulterior motives and conduct deceptive acts in ethical decision-making processes (Carlson et al., 2022; Perry, 2023). The potential benefits (e.g., wealth, fame) from concealing ill motives in our business ethics settings, which often make humans vulnerable to moral lapsing, do not appeal to AIs as these motives are irrelevant to them (Greene, 2014). Our findings corroborate that consumers do make inferences about this proposition, inferences that are consequential for consumer behavior, such as purchasing or not purchasing products and services from firms associated with the risk of such moral failures.

Our theorizing, notably, goes beyond mind perception as an explanation for how humans make sense of AI. Complementing the purely cognitivist account of mind perception that understands mind as a set of mental capabilities (Gray & Wegner, 2012), we take a more expansive stance, one that highlights the role of attribution, motivations and intentions when making inferences about what a mind is (Malle, 1999; Monroe & Malle, 2017). It is precisely through the lens of *motivational* mind inferences that we understand why it appears beneficial to use an AI in a moral compliance setting for an observer. Consumers make the inference that there is a fundamental motivational difference between humans and AI: While humans can and often do self-deal, it would be meaningless for an AI to do so. By identifying the perceived lack of ulterior motives as an attributional mechanism, we complement purely cognitivist accounts of AI minds, demonstrating that consumers' acceptance of AI as a moral compliance agent hinges not just on mental capabilities, but on attributions of intention and motivation – unlocking a deeper, theoretically rich understanding of how consumers make sense of AI in morally charged domains.

**Practical Implications**

Our study also offers practical implications. First, the findings suggest that as much as AI is often perceived as "moral threat," it can also be a seen in consumers' eyes as "moral good." It is unsurprising that people see AI as a moral threat in an agentic decision-making role. This feeling may result from violating traditional expectations about human exclusivity to moral decision-making (Bonnefon et al., 2016). The rapid development and adoption of AI may also contribute to such negative views. In many sectors, workers increasingly worry that AI will replace them, given its increasing ability to automate business processes (Davenport et al., 2020). In consumer products, AI is seen as a possible moral threat, associated with privacy breaches, manipulation of consumer behavior and perpetuation of biases (Crawford, 2024).

However, our study reveals insights that can balance the widespread narrative of AI as a moral threat. We demonstrate how AI can instead be viewed as morally positive by consumers, due to a perception that it lacks ulterior motives. Our study provides evidence relevant for managers grappling with AI ethics, demonstrating that giving AI a moral compliance role can bring both internal and external benefits. It can improve moral compliance in corporate governance (e.g., sticking to CSR goals and reporting and avoiding hypocrisy; Wagner et al., 2009) and, at the same time, generate positive consumer perceptions of the company and their products and services (an "AI halo effect").

Second, the study reveals that consumers have a more nuanced understanding of AI's moral potential than previously thought. While individuals may continue to believe that AI excels in functional domains rather than in subjective or emotional ones (Bigman & Gray, 2018; Castelo et al., 2019), rendering moral decision-making an ill fit; they seem to recognize that this cold impartiality can be an advantage when AI has a different business ethics *role*: one of dispassionately upholding norms (Bartels et al., 2015). This more nuanced understanding we capture should give managers comfort that consumer assessments of AI are not unidimensional, and not uniformly negative (Bowman, 2024).

Finally, our paper points to an opportunity for managers to reverse consumer AI aversion by deploying AIs in compliance roles, working in tandem with humans. This is especially relevant when aversion itself poses operational challenges. For example, social media platforms rely heavily on AI for content moderation and curation. Yet user dissatisfaction often arises when AI fails to handle morally ambiguous or conversely, straightforward cases effectively. Managers can mitigate this resistance by re-positioning AI not as a standalone decision-maker but as a tool that supports human decision-making while ensuring alignment with moral norms. This reframing not only addresses concerns about AI's limitations but also helps rebuild trust in its application. By clearly communicating AI's role as a partner in moral compliance, companies can foster greater acceptance and enhance their reputation for ethical operations.

**Limitations and Future Directions**

Further research is needed to model more complex forms of human-AI interactions in organizational settings and to test consumer reactions to such ensembles (Grote et al., in press; Raisch & Fomina, 2025; Choudhary et al., 2025). Our focus was a relatively simple base case to maximize theoretical clarity: AIs' singular role is not moral decision-making but moral compliance, their task is to oversee humans complying with ethical rules and norms, which have already been set (by other humans).

However, other scenarios are also possible, especially as AIs further develop. For example, as AIs may go beyond the constraints of mere upholding of preset norms by recognizing errors. An AI tasked with modern slavery compliance monitoring may recognize hard to detect non-compliance, which human agents may overlook (Mila_[Quebec_Artificial_Intelligence_Institute], 2025). By acting on such bottom-up, inductive insights AIs may start to move away from the role of mere moral compliance to segue into moral decision-making (by reinterpreting the confines of the moral norms they were set to police). If consumers have an understanding of such shifts in scope, they may reevaluate their moral assessments of AIs. Further, while our theorization clearly distinguishes moral compliance from moral decision-making (Geva, 2000, 2006; Julmi, 2024; Sonenshein, 2007), we acknowledge that compliance may still involve some discretion and context-sensitive judgment, particularly when rules are ambiguous or competing norms are in play. As our final study (Study 5) demonstrates, consumers clearly see a distinction – more research is needed in defining what moderating influences can increase of decrease differences defined here.

A related area needing serious investigation relates to agency theory (Jensen & Meckling, 1976). While outside the scope of the current investigation, consumers may have lay economic agency beliefs about principal-agent problems within the company deploying the AI. While many such beliefs are possible, one that suggests that strategic employees may turn AIs to act as their own agents (at the expense of the company and its principals) is particularly interesting. In such a scenario, the default consumer intuitions we captured (i.e., AIs lack ulterior motives) may change given a perception that a human may control the AI to extract unethical benefits. Consumers may perceive that human employees can be successful in "imputing" their own ulterior motives into AIs, they may believe AIs act as constant checks despite these attempts, or they may perceive a combination of these options, potentially bounded by a range of moderating features of the actors and their environment – with consequential outcomes for consumer evaluations. This area of conjoint human-AI agency and its perception among consumers has great promise as a research avenue as AIs gain more prominent roles in organizations.

Finally, future studies should investigate cultural variables as moderators. Prior work in the context of political corruption (Castelo, 2024) shows that people turn to AI when human decision makers are seen as susceptible to self-serving or corrupt motives. While this is a similar logic to our study, Castelo (2024) demonstrates the effects of perceived AI impartiality in different problem spaces and varying cultural contexts. We study AI as an enforcer of already-agreed moral rules and show that AI is preferred because it is seen as lacking ulterior motives and therefore more impartial than conflicted humans in compliance roles. Castelo (2024), by contrast, examines resource-allocation decisions and shows that in higher-corruption contexts people lean toward algorithms because they expect them to curb human corruption and to deliver productivity gains. Putting the two accounts together suggests that AI preference should be stronger in cultural or institutional contexts where (a) corruption is chronically salient, (b) trust in human gatekeepers is low, or (c) conflicts of interest are routine. This is because in all of these settings AI signals rule-bound, non-extractive enforcement. Future research could therefore test our compliance-based effect under experimentally heightened corruption salience or cross-national variation in institutional trust, to see whether the advantage we document for AI moral compliance agents is amplified in less democratic, more corrupt and higher power distance political environments across the globe.

In conclusion, our study breaks new ground in theorizing consumers' evolving understanding of AI morality, as deployed by marketing organizations – and opens important research avenues. As we demonstrate, consumers' understanding of AI morality is multidimensional. AIs may remain "unexplainable" for the foreseeable future, especially, in complex domains such as business ethics. Consumers, however, are already showcasing nuanced social explanations about AIs' moral roles on their own.

Wilson, S., & Haslam, N. (2009). Is the future more or less human? Differing views of humanness in the posthumanism debate. *Journal for the Theory of Social Behaviour*, *39*(2), 247–266. https://doi.org/10.1111/j.1468-5914.2009.00398.x

World_Bank. (2020, March 18). *Artificial intelligence in the public sector : Maximizing opportunities, managing risks (vol. 1 of 2)*. Retrieved November 10, 2025 from http://documents.worldbank.org/curated/en/809611616042736565

Xu, X., Xiong, F., & An, Z. (2023). Using machine learning to predict corporate fraud: Evidence based on the gone framework. *Journal of Business Ethics*, *186*(1), 137–158. https://doi.org/10.1007/s10551-022-05120-2

Yang, L. W., Aggarwal, P., & McGill, A. L. (2020). The 3 c's of anthropomorphism: Connection, comprehension, and competition. *Consumer Psychology Review*, *3*(1), 3–19. https://doi.org/10.1002/arcp.1054